\documentclass[12pt]{iopart}

\usepackage{graphicx}
\usepackage[table]{xcolor} 
\usepackage{booktabs} 
\usepackage{upgreek}
\usepackage{caption}

\begin{document}
\title[Optical Inspection of European XFEL Cavities]{Optical Surface Properties and their RF Limitations of European XFEL Cavities}
\author{Marc Wenskat}
\address{Deutsches Elektronen Synchrotron, 22607 Hamburg, Notkestrasse 85}
\ead{marc.wenskat@desy.de}
\vspace{10pt}
\begin{indented}
\item[]March 2017
\end{indented}

\begin{abstract}
The inner surface of superconducting cavities plays a crucial role to achieve highest accelerating fields and low losses. The industrial fabrication of cavities for the European X-Ray Free Electron Laser (XFEL) and the International Linear Collider (ILC) HiGrade Research Project allowed for an investigation of this interplay. For the serial inspection of the inner surface, the optical inspection robot OBACHT was constructed and to analyze the large amount of data, represented in the images of the inner surface, an image processing and analysis code was developed and new variables to describe the cavity surface were obtained. This quantitative analysis identified vendor specific surface properties which allow to perform a quality control and assurance during the production. In addition, a strong negative correlation of $\rho= -0.93$ with a significance of $6\,\upsigma$ of the integrated grain boundary area $\sum{\mathrm{A}}$ versus the maximal achievable accelerating field $\mathrm{E_{acc,max}}$ has been found. 
\end{abstract}

\vspace{2pc}

\section{Introduction}
Superconducting niobium radio-frequency cavities are fundamental for the European XFEL and the ILC \cite{XFEL_TDR,ILC_TDR}. To utilize the operational advantages of superconducting cavities, the inner surface has to fulfill quite demanding requirements. 
The surface roughness, welding techniques, and cleanliness improved over the last decades and with them, the achieved maximal accelerating gradient $\mathrm{E_{acc,max}}$. Still, limitations of the accelerating gradient are observed, which are not explained by localized geometrical defects or impurities. The method and results shown in this paper aim for a better understanding of these limitations in defect free cavities based on global, rather than local, surface properties.
 
\section{Optical Inspection and Image Processing}
For this research, more than 100 cavities underwent subsequent surface treatments, cold RF tests, and optical inspections within the ILC-HiGrade research program and the European XFEL cavity production \cite{CORDIS, Navitski2013,Singer2016}. The optical inspection of the inner surface of superconducting RF cavities is a well-established tool at many laboratories \cite{Watanabe}. 
Its purpose is to characterize and understand field limitations and to allow optical quality assurance during cavity production. Theoretical calculations in \cite{Sara1995,Xie2009} have shown that accelerating fields of $50\, \mathrm{MV/m}$  are achievable if surface structures and localized defects are below 10\,$\upmu \mathrm{m}$, hence the resolution of the system should be on that order. 

An algorithm was developed which enables an automated surface characterization based on images taken by an optical inspection robot. This algorithm delivers a set of optical surface properties, which describe the inner cavity surface. These optical surface properties deliver a framework for a quality assurance of the fabrication procedures. Furthermore, they show promising results for a better understanding of the observed limitations in defect free cavities.
\subsection{OBACHT}
\begin{figure}[!htb]
	\centering
		\includegraphics[width=0.75\textwidth]{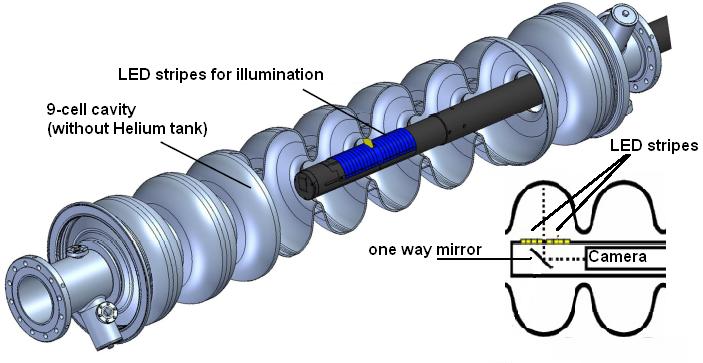}
	\caption{Schematic of the Kyoto Camera System used at DESY. The camera is viewing the inner surface via a $ 45^\mathrm{o}$-tilted one way mirror. Behind this one way mirror, three LEDs are mounted for the central illumination. $2 \times 10$ acrylic strips with LEDs are mounted left and right from the opening in the tube for a more detailed illumination. }
	\label{fig:sketch}
\end{figure}

A fully automated robot for optical inspection,  the "\textbf{O}ptical \textbf{B}ench for \textbf{A}utomated \textbf{C}avity inspection with \textbf{H}igh resolution on short \textbf{T}imescales" (OBACHT), has been continuously used at DESY. It is equipped with a high-resolution camera (Kyoto Camera System), which resolves structures down to $12\,\upmu \mathrm{m}$ for properly illuminated surfaces \cite{Iwashita2008,Tajima2008,Iwashita2009}. The details of OBACHT and the optical system are described in \cite{Lemke,Sebastian,Wenskat2015}. It consists of a camera tube with a diameter of 50\,mm to fit into the cavity without colliding with the irides or HOM antennas protruding into the cavity volume (see Fig. \ref{fig:sketch}). In this tube the camera, together with a low-distortion lens, are installed. The camera system images the surface via a $ 45^\mathrm{o}$-tilted one way mirror which can be continuously adjusted to other angles in order to inspect other cavity regions. The focal distance of the camera to the cavity surface is controlled by a motor driven lead-screw. For illumination, acrylic strips (two LEDs per strip) attached to the camera tube around the camera opening are installed, together with three additional LEDs behind the one way mirror inside the camera tube.

\begin{figure}[!htb]
	\centering
		\includegraphics[width=0.50\textwidth]{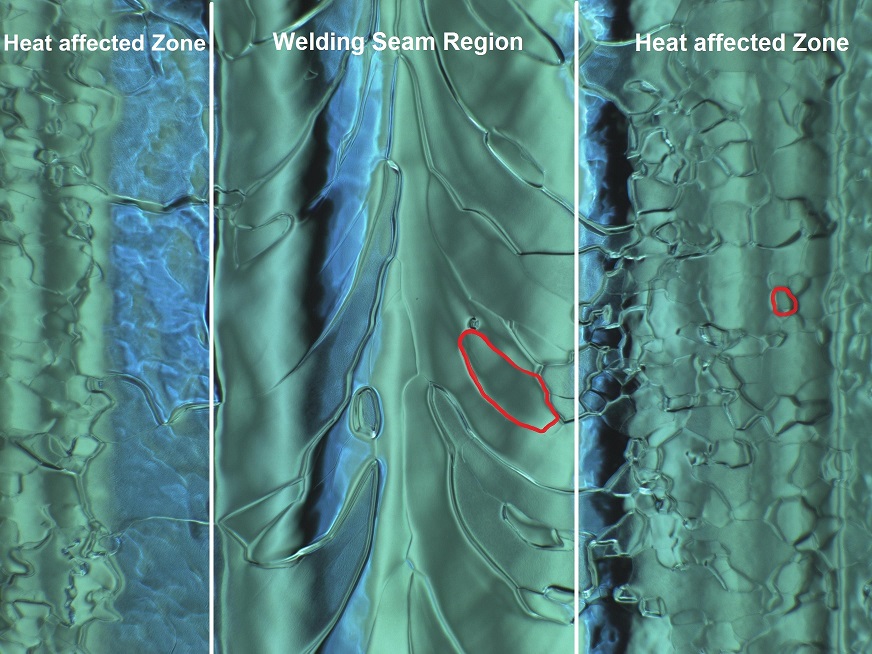}
	\caption{Image of the inner cavity surface with the equatorial welding seam in the image center taken with OBACHT. The image size is $9 \times 12\,\mathrm{mm^2}$. The red contours are examples of grain boundaries identified with the image processing algorithm in the welding seam region (WS) and the heat affected zone (HAZ).}
	\label{OBACHT}
\end{figure}

The highest magnetic field in a cavity, and hence the highest losses, are at the equatorial welding seam region including the heat affected zone. Therefore, the equatorial images are of main interest for this analysis. Figure \ref{OBACHT} shows an image of the inner cavity surface taken with OBACHT.
With given cavity geometry and optical set up, an individual image covers $5^\mathrm{o}$ of an equator. To have a small overlap at the edges of an image, an image is taken each $4.8^\mathrm{o}$. This results in 75 images per equator and 675 equator images per cavity. The objects of interest within an image of the inner cavity surface are grain boundaries. In order to identify and quantify those boundaries, an image processing and analysis algorithm has been developed.

\subsection{Image Processing and Analysis}
\begin{figure}[!htb]
   \centering
   \includegraphics[width=0.60\textwidth]{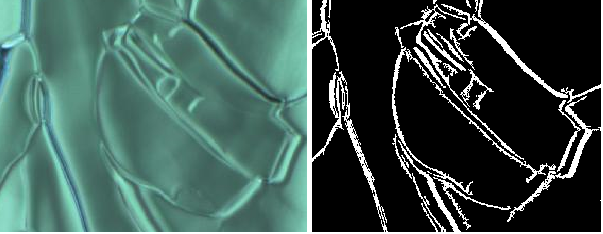}
   \caption{Left: a detail of an OBACHT image is shown. Right: the same detail after the image processing algorithm. Grain boundaries (white) are visible.}
   \label{GB_area}
\end{figure}

The main goal of the image processing algorithm is to identify grain boundaries, regardless of their position within the image which shows a non-uniform illumination, as can be seen in Figure \ref{OBACHT}. The approach of this algorithm is, to apply a sequence of high-pass filter and local contrast enhancements, to project pixels which belong to grain boundaries onto a gray scale interval, which is distinct to the background. After this projection, a histogram based segmentation of the processed image is performed. This segmentation assumes, that the image contains two classes of pixels (grain boundary and background), where the intensity values follow a bi-modal distribution, and calculates the optimum threshold separating the two classes. 
The output is a binary image with the same size as the input image. It will contain grain boundary pixels in white (logical one) and background pixels in black (logical zero). As a last step of the image processing, groups of connected white pixels which form a grain boundary have to be classified as a single object and a labeled binary image is obtained. 
An example of such a binary image is given in Figure \ref{GB_area}.
\newline
The aim of the image analysis is to identify features in the binary image. Those are grain boundaries with varying width which are not symmetric. Hence, it is nontrivial to define important properties like diameter, centroid, eccentricity or orientation of an object. The method to overcome this problem, is to find an ellipse which has the same second central moment as the pixel distribution of the pixel \cite{Hu1962}. Within this framework, the grain boundary area is the total amount of pixels, consisting of a boundary. This number is retrieved from the binary image and then multiplied by the pixel size, which is a property of the optical system. At OBACHT, this value is $12.25\,\upmu \mathrm{m}^2$. With the given resolution at OBACHT, the experimentally obtained relative error for the grain boundary area is 3\,\%, similar to \cite{Patil2011}. 

The orientation of an object is defined with respect to an axis perpendicular to the welding seam and an upper limit of the uncertainty of $ 5^\mathrm{o}$ is derived with \cite{Klette1999,Liao1993}. 

In order to define a figure of merit for the roughness of an object with OBACHT, two assumptions were made. The first assumption is that the intensity of the reflected light is dependent on the roughness and structures of the cavity surface. This means that a change in the intensity is either caused by a geometric gradient or a change in reflectivity. A geometric gradient exists either at a grain boundary or a defect, while a change in reflectivity can be caused by an impurity. The second assumption is that the curvature of the elliptical cavity is negligible within the studied area and the surface seen by the image can be considered to be a flat surface. 
Based on the intensity of the original image, a quantity called $\mathrm{R_{dq}}$ is introduced. It is based on ISO 25178 for surface texture \cite{iso25178} and is the average of the intensity gradient of the boundary. A steeper slope of an edge or surface would imply a larger intensity gradient and hence a larger $\mathrm{R_{dq}}$. 

A statistical noise arising from the Signal-to-Noise-Ratio (SNR) of the image sensor in the camera, which yields a $\delta \mathrm{R_{dq}}$ of $\frac{0.011}{\sqrt{N}} \frac{\mathrm{Bit}}{\mathrm{\upmu m}}$. A systematic uncertainty due to image focus was found to be  $\frac{\delta \mathrm{R_{dq}}}{\mathrm{R_{dq}}}= 3\,\%$. For more details on the image processing algorithm and explicit definitions and discussion of the obtained variables see \cite{Wenskat2015}.

\section{Vendor dependent surface properties}
As described in \cite{Singer2016, Aderhold2010}, the two cavity vendors, RI Research Instruments GmbH (RI) and Ettore Zanon S.p.A. (EZ), were qualified to produce cavities after two distinct procedures. Most notably is the electron beam welding (EBW) procedure as well as the final surface chemistry step since both have a significant influence on the final surface and therefore RF performance. Since only the standard fabrication procedure should be compared, cavities that underwent any repair were not used in this comparison, which reduced the usable data set since most of the inspected cavities were issued an inspection due to flaws in the fabrication. This data set than yields to a total number of nine RI and eleven EZ cavities which were inspected with OBACHT.   
    
\subsection{Electron Beam Welding Procedure}
\begin{figure}[!htb]
	\centering
\includegraphics[width=1\textwidth]{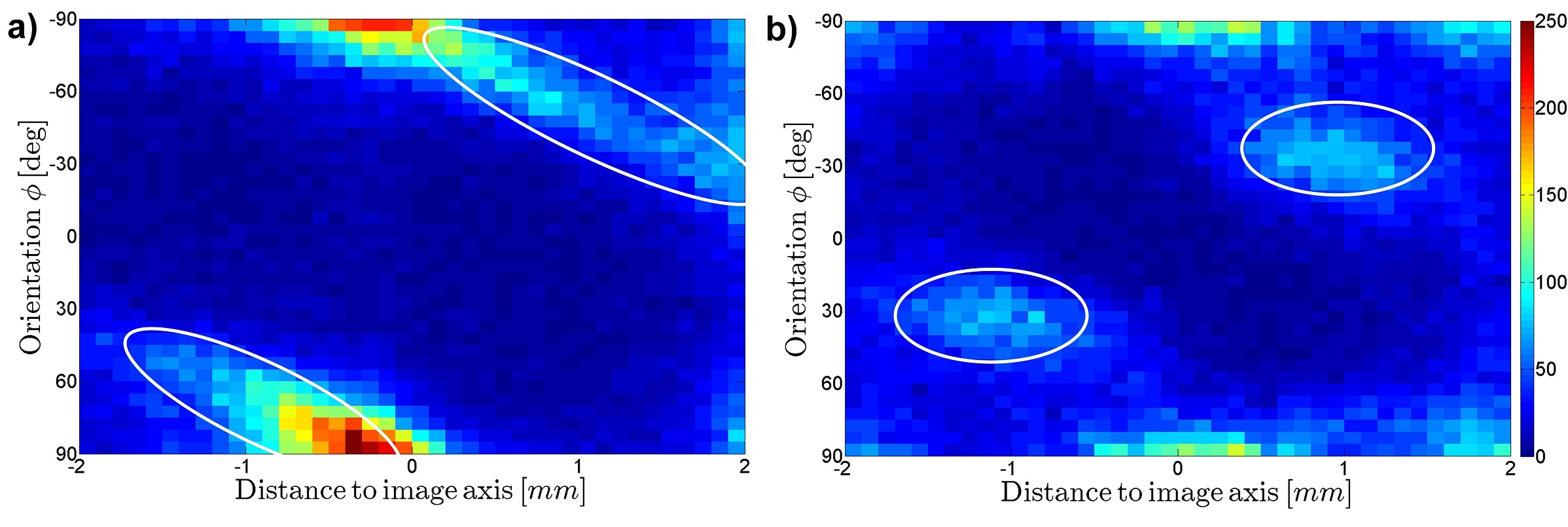}
	\caption{The x-axis shows the boundary centroid position with respect to the image mirror axis, which is the welding seam ridge. The y-axis shows the boundary orientation $\phi$ in degrees w.r.t. an axis perpendicular to the welding seam. Only the welding seam region is shown. The left plot (a) represents boundaries in the welding seam region of a RI cavity, the right plot (b) of an EZ cavity. The white ellipses encircles the welding seam boundaries. The color depicts the counts per bin.}
	\label{Orienthist_vendors}
\end{figure}

The solidification dynamics of the weld puddle depends on parameters such as temperature gradient, crystalline growth rate, and chemical composition. Therefore, the granular microstructure which develops in the molten material varies and depends on the weld movement pattern, beam travel speed, and beam power.
The histograms in Figure \ref{Orienthist_vendors} show the grain orientation in the welding seam region as obtained with the image processing algorithm for each vendor. 
\newline
As it can be seen, the boundaries in the welding seam region of EZ have angles of $ \pm\,30^\mathrm{o}$. The angles of the boundaries in the welding seam region of a RI cavity show a completely different distribution. At the edge of the welding seam, the boundaries have an angle of $ \pm\,30^\mathrm{o}$, similar to EZ, while the boundaries change their orientation towards the center of the welding seam. 

\begin{figure}[htbp]
	\centering
    \includegraphics[width=0.68\textwidth]{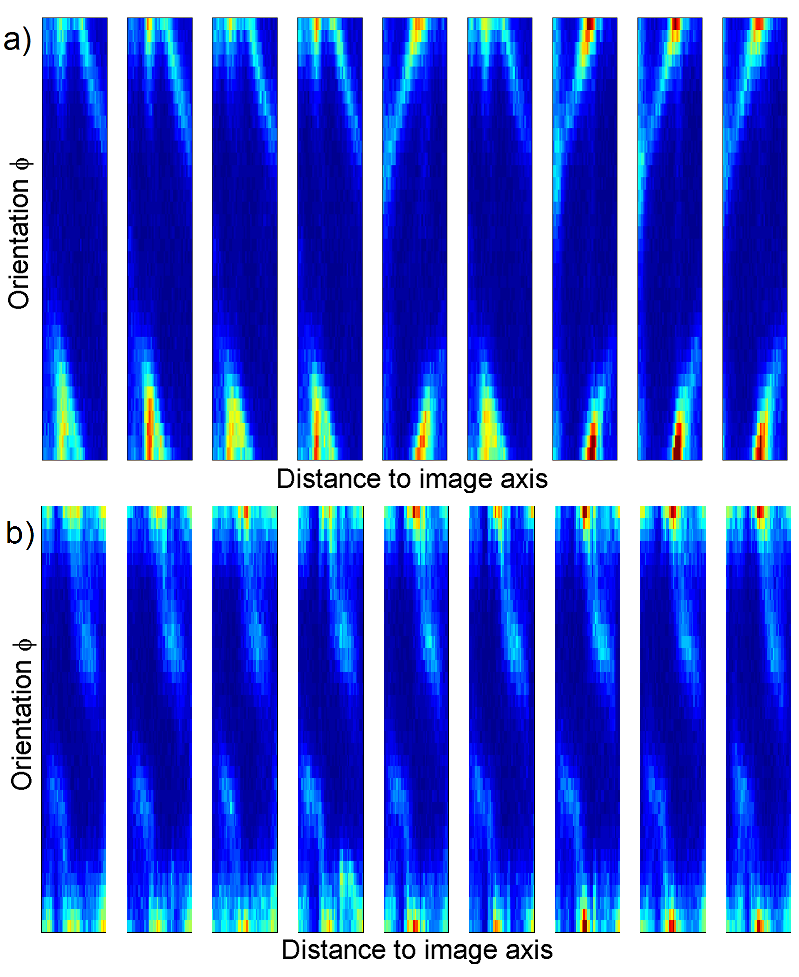}
	\caption{Within the histograms, the color depicts the counts per bin. The x-axis shows the boundary centroid position with respect to the image axis. The y-axis shows the boundary orientation $\phi$ in degrees w.r.t. an axis perpendicular to the welding seam. Only the welding seam region is shown. a) is for RI and b) for EZ. Each histogram represents a different equator along the cavity - equator 1 is on the left.}
	\label{Orienthist_vendors_patterns}
\end{figure}

Another observed difference between the vendors is shown in Figure \ref{Orienthist_vendors_patterns}.

All equator orientation patterns for EZ point into the same direction while they flip between equator 4-5, 5-6, and 6-7 for RI.

\subsection{Surface Roughness}
\begin{figure}[!htb]
	\centering
		\includegraphics[width=0.90\textwidth]{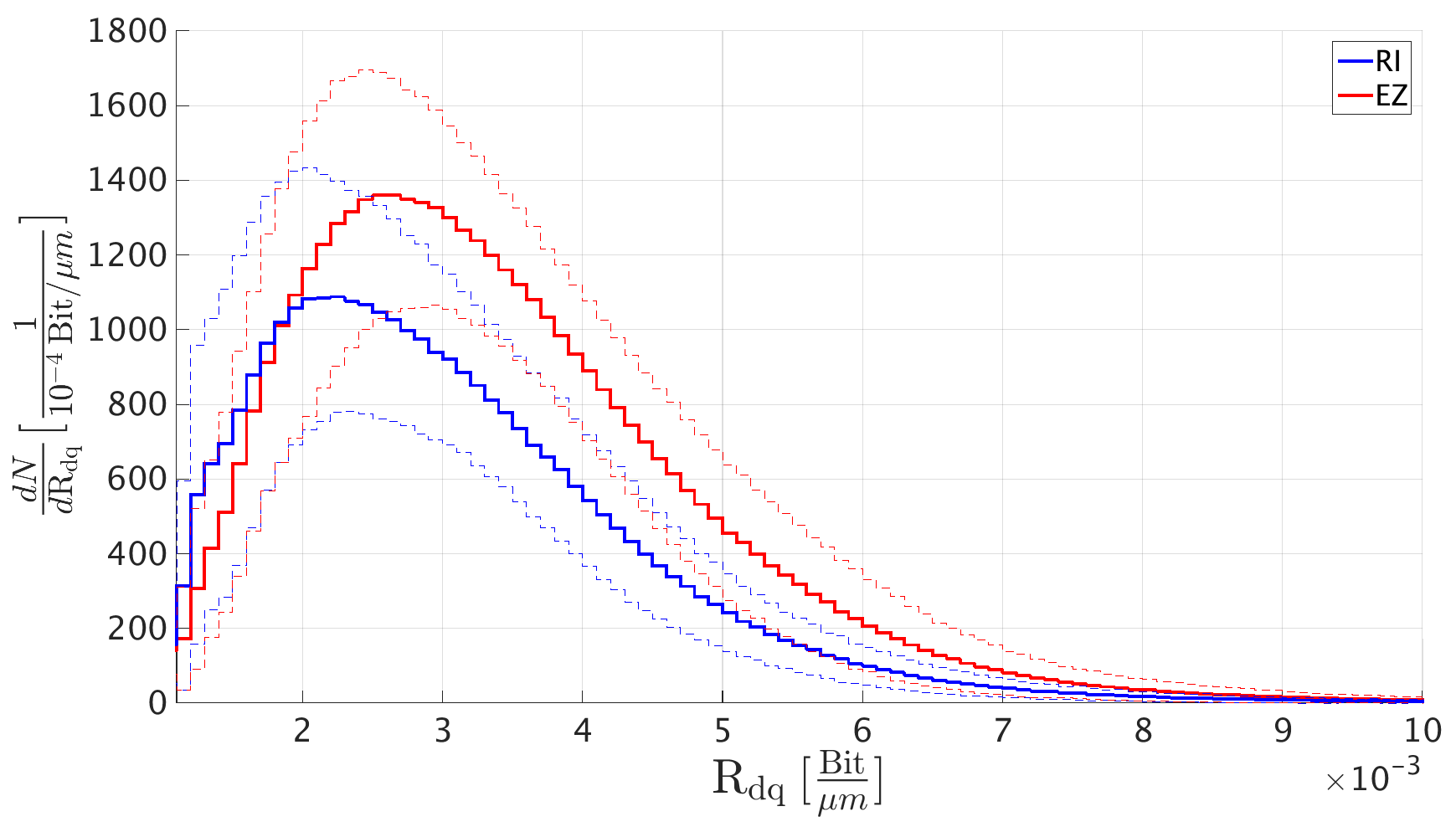}
	\caption{The x-axis shows the grain boundary gradient $\mathrm{R_{dq}}$ for boundaries within the welding seam region and the y-axis shows the counts per bin. The plots show the average $\mathrm{R_{dq}}$ distribution for all cells with a one $\upsigma$ interval. The red distribution is for EZ (N=99) and the blue distribution is for RI (N=81).}
	\label{Vendor_Rdq_Distribution}
\end{figure}

Cavities from both vendors underwent a bulk electro-polishing procedure (EP). In addition, the cavities produced by RI underwent a final electro-polishing procedure (EP) of 40\,$\upmu \mathrm{m}$ while the EZ cavities underwent a flash buffered chemical polishing (BCP) of 10\,$\upmu \mathrm{m}$. The difference between these surfaces, as parametrized by $\mathrm{R_{dq}}$, is shown in Figure \ref{Vendor_Rdq_Distribution}.
\newline
For a quantification, an exponential modified Gaussian distribution (EMG) is fitted to the observed distribution. The fit parameters are given in Table \ref{tab:Rdq_emg_fit_vendor} and the average value of the distribution in Table \ref{tab:Rdq_emg_fit_vendor_mlv}.
\begin{table}[ht]
\small
	\centering
\makebox[\linewidth]{
		\begin{tabular}{l>{\centering\arraybackslash} p{1.5cm}>{\centering\arraybackslash}  p{1.5cm}>{\centering\arraybackslash}  p{1.5cm}>{\centering\arraybackslash}  p{1.5cm}>{\centering\arraybackslash}  p{1.5cm}>{\centering\arraybackslash}  p{1.5cm}}
 \toprule
 \multicolumn{1}{c}{ }           &    \multicolumn{3}{c}{\textbf{\normalsize{WS}} } &    \multicolumn{3}{c}{\textbf{\normalsize{HAZ}} }    \\
		
		 $\left[ 10^{-3} \frac{\mathrm{Bit}}{\upmu \mathrm{m}}\right]$           &  $ \upmu$    & $\upsigma $ & $\uplambda^{-1}$ &  $\upmu$ & $\upsigma $ & $\uplambda^{-1}$ \\
		             \cmidrule{2-7}
		\textbf{\normalsize{RI}}   &	$1.6 \pm 0.1 $ & $0.7 \pm 0.1$ & $1.2 \pm 0.1$ & $2.0 \pm 0.1$ & $0.9 \pm 0.2$  & $1.1 \pm 0.1$  \\ 
		\textbf{\normalsize{EZ}}    &	$2.2 \pm 0.1 $ & $0.8 \pm 0.1$ & $1.2 \pm 0.1$ & $2.3 \pm 0.1$ & $0.8 \pm 0.2$  & $1.3 \pm 0.1$  \\ 
		\bottomrule
		\end{tabular}
		}
			\caption{The fit parameters $\upmu$ is the mean, $\upsigma $ the variance of the Gaussian component and $\uplambda^{-1}$ the inverse exponential decay rate of the EMG distribution, in the welding seam region (WS) and the heat affected zone (HAZ) with the 95\% confidence interval.}
		\label{tab:Rdq_emg_fit_vendor}
\end{table}
\begin{table}[ht]
	\centering
		\begin{tabular}{l>{\centering\arraybackslash} p{1.8cm}>{\centering\arraybackslash} p{1.8cm}}
		\toprule
		\multicolumn{1}{c}{ }           &    \multicolumn{2}{c}{\textbf{\normalsize{$\overline{\mathrm{R_{dq}}}$}} }   \\
		$\left[ 10^{-3} \frac{\mathrm{Bit}}{\upmu \mathrm{m}}\right]$  & \textbf{WS} & \textbf{HAZ}    \\ 
		\cmidrule{2-3}
		\textbf{RI}  & $2.8 \pm 0.1  $  & $3.1 \pm 0.1 $  \\ 
		\textbf{EZ}  & $3.4 \pm 0.1  $  & $3.6 \pm 0.1 $  \\ 
	  \bottomrule			
		\end{tabular}
			\caption{The average $\mathrm{R_{dq}}$ derived by the EMG fit with 95\% confidence interval.}
		\label{tab:Rdq_emg_fit_vendor_mlv}
\end{table}
\newpage
As seen in the histograms and quantified by the values of the EMG parameters, the cavities produced by RI have, on average, a smaller $\overline{\mathrm{R_{dq}}}$ of 17\,\% in the welding seam (WS) region in comparison to cavities produced by EZ. 
The biggest difference of the average roughness, quantified by $\upmu$, is found in the welding seam region. The heat affected zone (HAZ) of BCP cavities show a slightly larger amount of steep grain boundaries, quantified by $\uplambda^{-1}$.

\section{Surface properties and RF performance}
The main motivation for the construction of an optical inspection for cavities was to gain a better understanding of their RF performance, respectively the quality factor $Q_0$ and the accelerating gradient $\mathrm{E_{acc,max}}$. Correlations between quenches or field emission during cold RF tests and localized defects seen in optical inspections are well known \cite{Watanabe,Sebastian,Moller2009,Geng2009a,Aderhold2010b,Singer2010}. A systematic study on the correlation of global optical surface properties and RF performances has not been performed yet.  

For a quantitative statement on the goodness of the relation between the obtained optical surface properties and a figure of merit of the RF performance, the Pearson correlation coefficient $\rho$ is used. Since both variables are subject to measurement uncertainties, the calculation of the correlation coefficient should include these uncertainties. With appropriate estimators, the corrected Pearson correlation coefficient can be deduced which includes the influence of the uncertainty on the correlation coefficient\cite{Pearson1966,Darmstadt2011}. The 95\,\% confidence interval of the correlation coefficient is calculated with a bootstrapping method \cite{efron1979,efron1987}. 
\subsection{Optical Surface Properties}
The purpose of this investigation is to identify the RF limiting cell. Two assumptions were made for this analysis. Firstly, that the maximal accelerating field $\mathrm{E_{acc,max}}$ shows a negative correlation to an optical determined variable. Secondly, the optically worst cell (aka the cell with the maximum value of this variable) should also be the RF limiting cell, since one bad performing cell is sufficient for a cavity to show a bad performance. Hence, the maximum value of the yet to identify optical surface variable of the nine cells identifies the \textit{optically conspicuous cell}. This maximum value is used to represent the whole cavity.
\newline
The different loss models discussed in the literature predict different correlations between the RF performance and the surface properties, but mainly all of them see the grain boundaries as a potential source for limitations or losses \cite{Visentin2003,Bauer2004,Ciovati2007,Ciovati2008a,hylton1988,Safa1999,Knobloch1999a}. Hence a property called \textit{integrated grain boundary area} $\sum{\mathrm{A}}$ in a cell as optical surface variable is used as a correlator within this work. This property is the sum of all grain boundary areas found in the 75 images of an equator.

\subsection{Correlation with 2nd Sound Results}
\begin{figure}[!htb]
	\centering
		\includegraphics[width=0.90\textwidth]{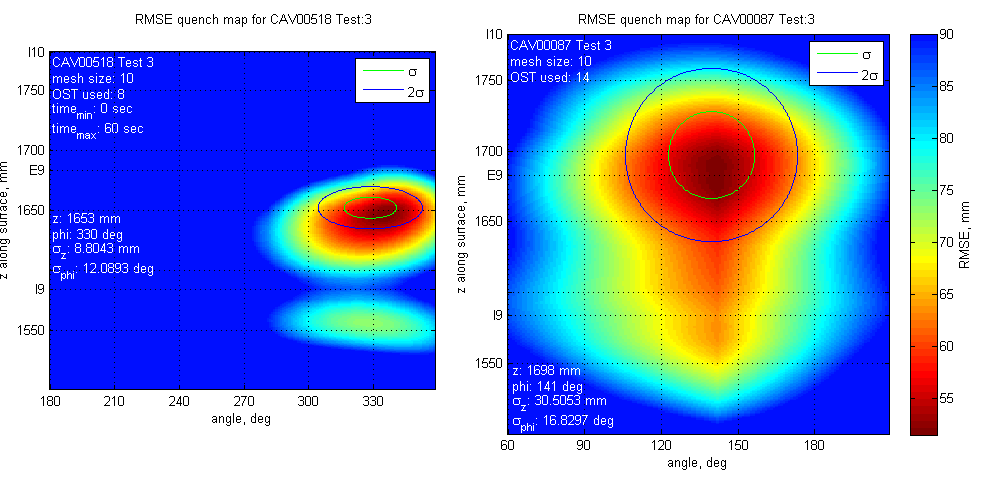}
	\caption{RMSE-Map of CAV00518 (left) and CAV00087 (right). The x-axis is the cell angle, the y-axis the longitudinal position along the cavity. The z-axis shows the color coded RMSE value. The most likely quench spot for CAV000518 is about 44\,mm below equator nine at ($330^\mathrm{o}\,\pm\,12^\mathrm{o}$) and for CAV00087 at equator nine at ($141^\mathrm{o}\,\pm\,16^\mathrm{o}$) \cite{Yegor2017}.}
	\label{fig:RMSEMap_Yegor}
\end{figure}

To test the assumptions, that the \textit{integrated grain boundary area} $\sum{\mathrm{A}}$ can be used as a correlator, a correlation of the quench location of a cavity and the image analysis is done, where a quench is the localized origin of the phase transition from the super- into the normal-conducting phase and limits the maximal accelerating field $\mathrm{E_{acc,max}}$. Only two of the cavities in the data set, CAV00518 and CAV00087, have been tested with the 2nd sound set up (\cite{2nd1, 2nd2}) and had a subsequent optical inspection . The quench location is obtained with a ray tracing method, described in \cite{Yegor2017}. The spot which minimizes the root-mean-square error (RMSE) between the theoretical and the experimentally 2nd sound velocity is the most likely origin of the 2nd sound wave and therefore the quench spot. The corresponding RMSE-maps for the two cavities are shown in figure \ref{fig:RMSEMap_Yegor}. 
\newpage
For both cavities, CAV0087 and CAV00518, the quench spot localized by 2nd sound was in cell nine. A visual inspection of the cavity surfaces using the 2nd sound system results as guidance did not reveal any local defect which could be identified as the origin of the quench, hence global surface properties are assumed to be the cause. The same equators were identified as optically conspicuous cells by the image analysis algorithm, as they had the largest \textit{integrated grain boundary area} $\sum{\mathrm{A}}$ of their cavities. The probability for this observation to be a coincidence is 1.2\,\%. 

\subsection{Optical Assessment of Cells}
\begin{figure}[!htb]
   \centering
   \vspace{5mm}
   \includegraphics[width=0.90\textwidth]{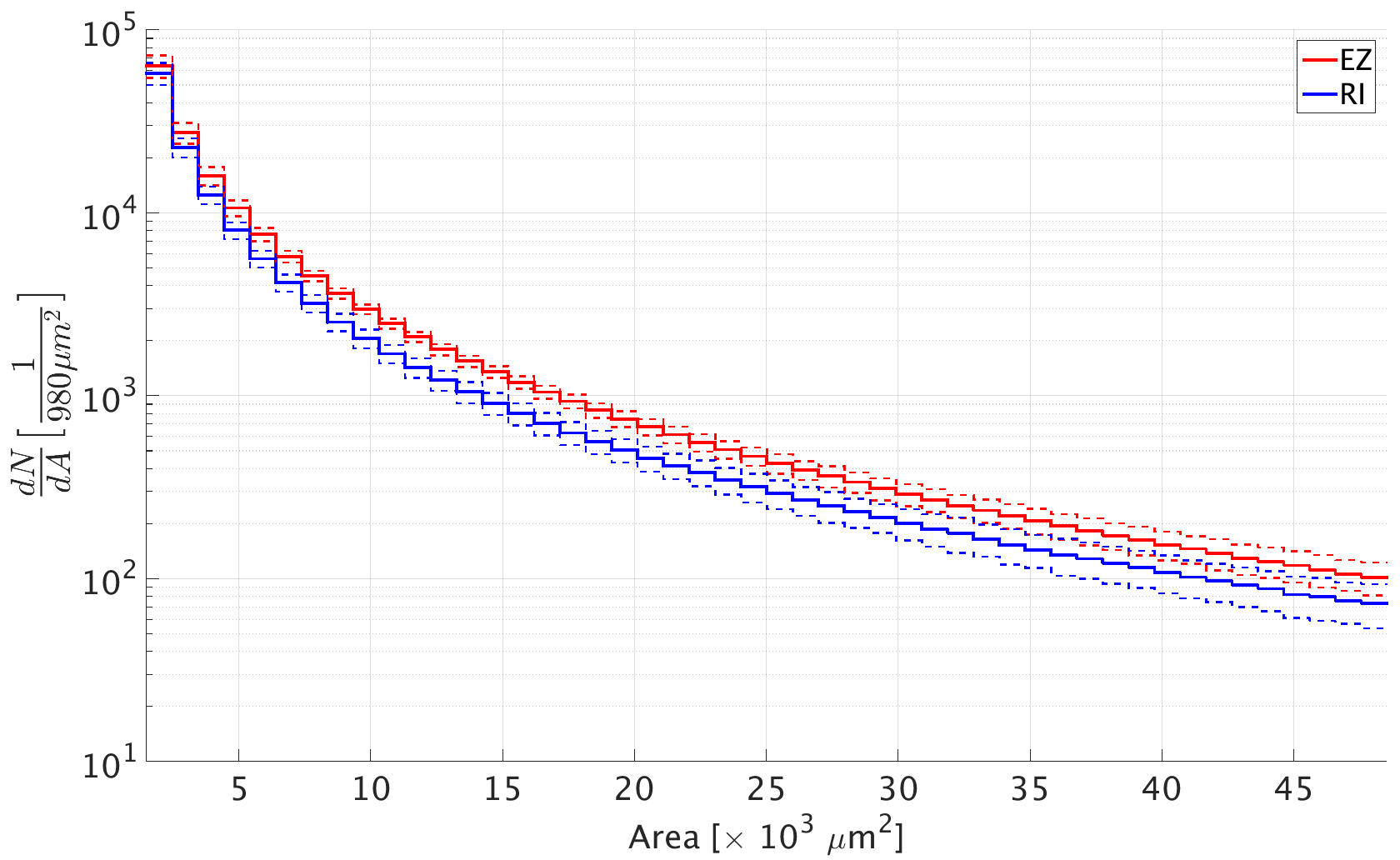}
   \caption{Average grain boundary area distribution of an equator. The blue distribution shows the average of 126 RI equators, the red distribution shows the average of 279 EZ equators. The dotted lines are the 1$\,\upsigma$ confidence intervals of the average.}
   \label{GB_area_distribution}
\end{figure}

\begin{figure}[!htb]
   \centering
   \vspace{5mm}
   \includegraphics*[width=0.70\textwidth]{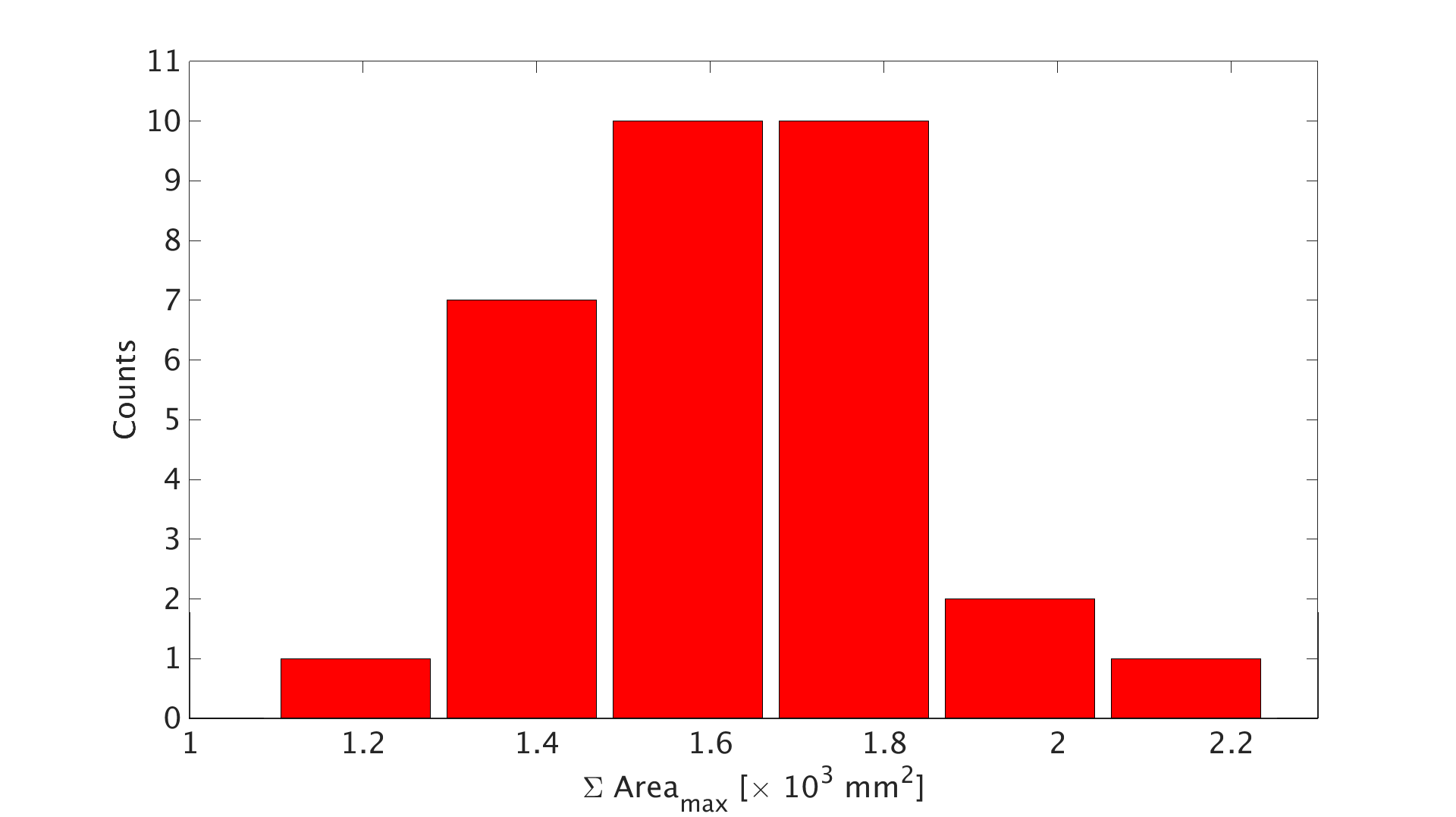}
   \caption{Histogram of the values of the integrated grain boundary area for the 31 optically conspicuous cells from EZ.}
   \label{Hist_EZ}
\end{figure}

\begin{figure}[!htb]
   \centering
   \vspace{5mm}
   \includegraphics*[width=0.80\textwidth]{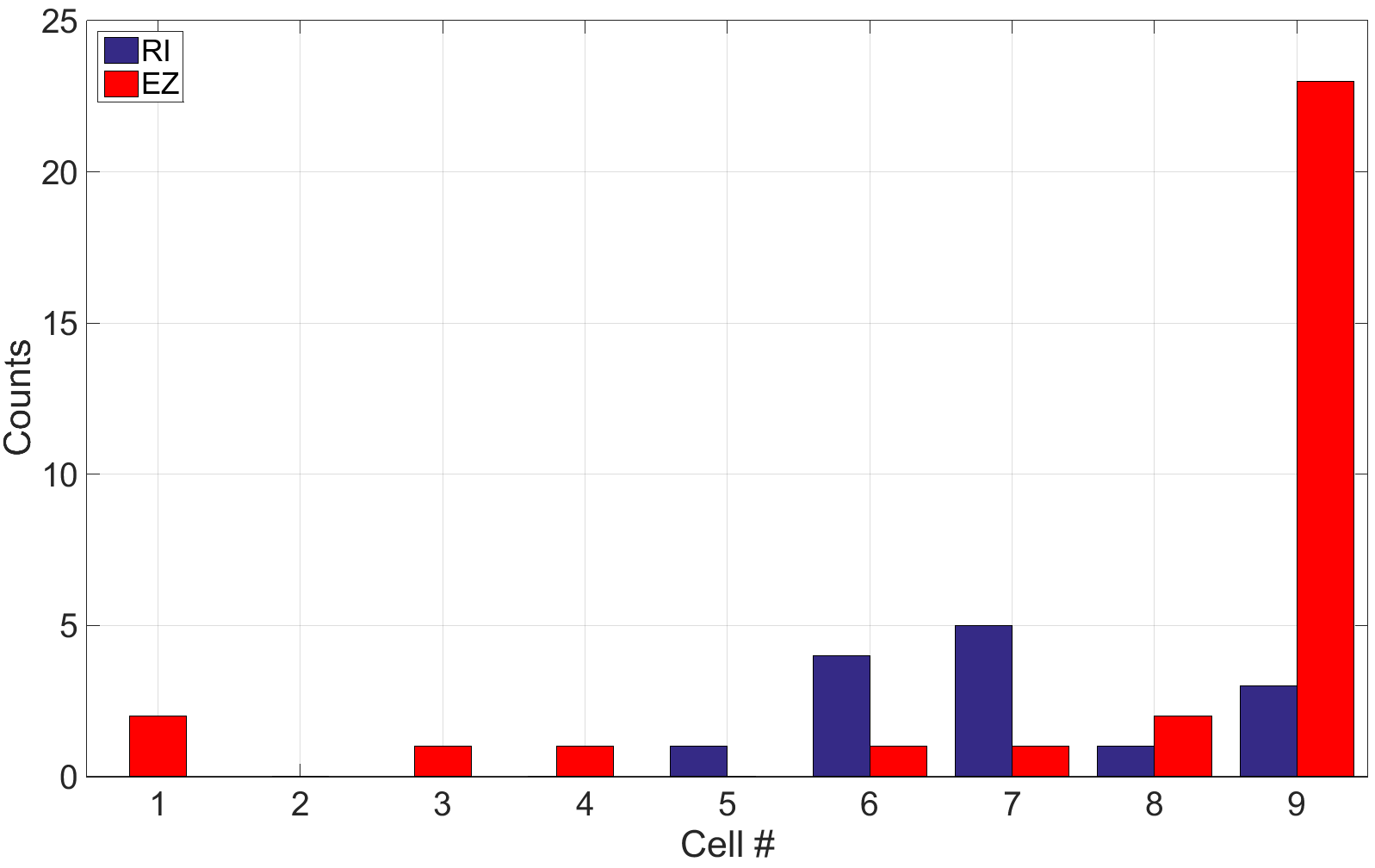}
   \caption{Observed longitudinal distribution of the \textit{optically conspicuous cells}. The blue distribution is derived from 14 RI cavities, the red distribution from 31 EZ cavities.}
   \label{Vendor_worst}
\end{figure}

\begin{figure}[!htb]
   \centering
   \vspace{5mm}
   \includegraphics*[width=0.80\textwidth]{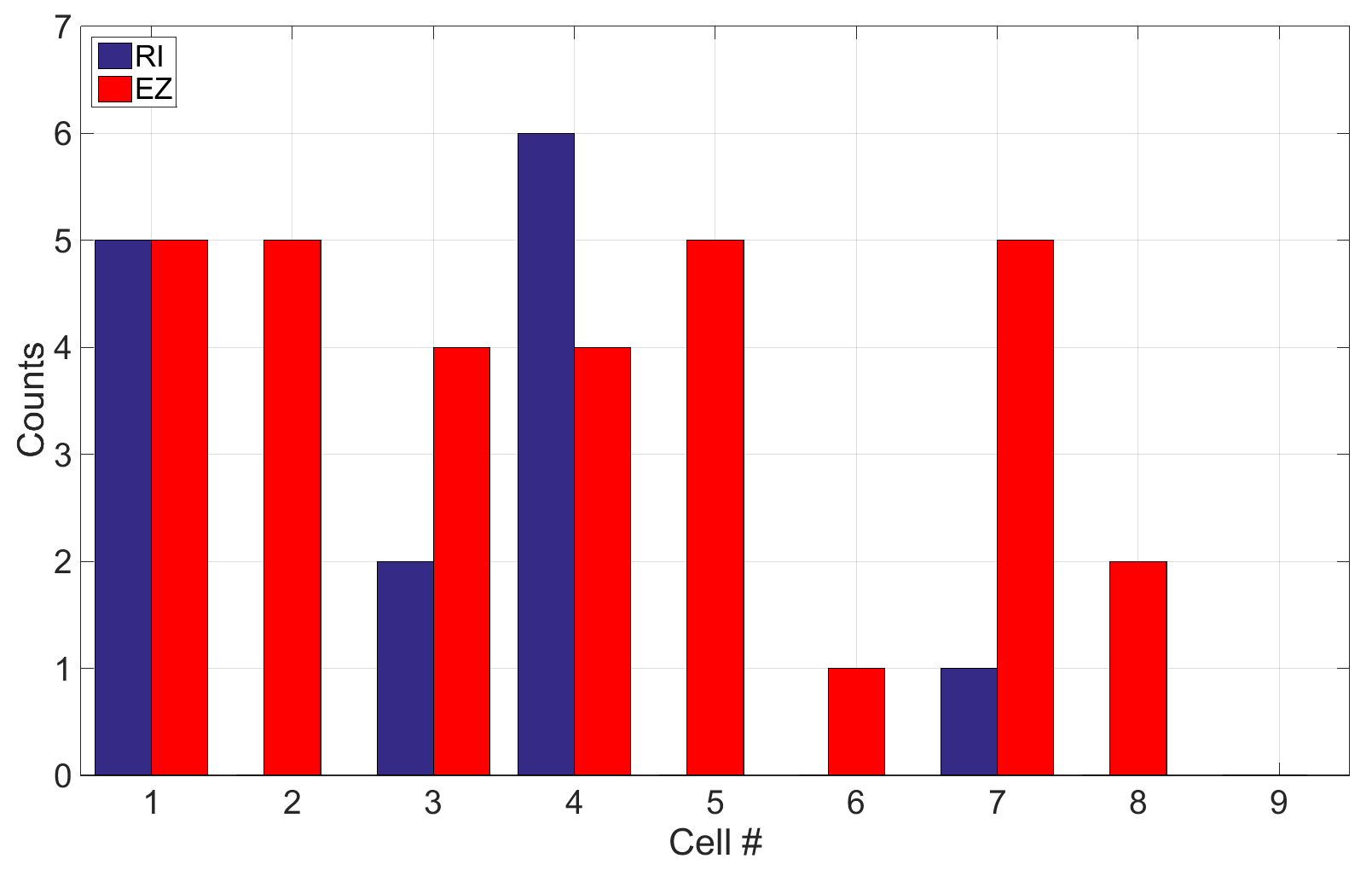}
   \caption{Observed longitudinal distribution of the \textit{optically best cells}. The blue distribution is derived from 14 RI cavities, the red distribution from 31 EZ cavities.}
   \label{Vendor_best}
\end{figure}

At the time of this analysis, 14 cavities from RI and 31 cavities from EZ were inspected. The observed grain boundary area distribution per equator is shown in figure \ref{GB_area_distribution}.
\newline
It can be seen that RI cavities exhibit a smaller number of boundaries with an area above $5 \cdot 10^3~\upmu \mathrm{m^2}$ than EZ cavities. 
Figure \ref{Hist_EZ} shows the histogram of the observed values of the integrated grain boundary area of the optically conspicuous cells for 31 cavities from EZ.
The integrated grain boundary area for EZ averages at $1600\,\pm 400\,\mathrm{mm^2}$, hence covering about $23\,\pm\,6$\,\% of the weld area. For RI, the average is $1200\,\pm\,260\,\mathrm{mm^2}$ and the coverage rate is $17 \pm\,4$\,\%. A priori, the longitudinal distribution of the optically conspicuous cell within a cavity should be a uniform distribution. The observed distribution of the optically conspicuous cells is shown in Figure \ref{Vendor_worst}.
\newpage
For EZ, equator 9 is the conspicuous cell in 24 out of 31 cases. For RI, equators 6 and 7 are the conspicuous cells in 9 out of 14 cases. The  $\chi^2$-test for both vendors showed a statistically significant deviation of the observed distribution from a uniform distribution with a $\chi^2$ of 17 for RI and a $\chi^2$ of 112 for EZ and the degrees of freedom (df) of 8 for both. 
The cell with the smallest integrated grain boundary area of a individual cavity is called the \textit{optically best cell}. The distribution of the optically best cells within the cavity is shown in Figure \ref{Vendor_best}.
\newpage
For EZ, the $\chi^2$-test showed that the observed distribution is in agreement with a uniform distribution with a $\chi^2$ of 8.2 and dof equal to 8.
For RI, equators 1 and 4 are favored. The $\chi^2$-test for a uniform distribution yields $\chi^2$ of 32 and dof equal to 8.

\begin{figure}[!htb]
   \centering
   \vspace{5mm}
   \includegraphics*[width=0.80\textwidth]{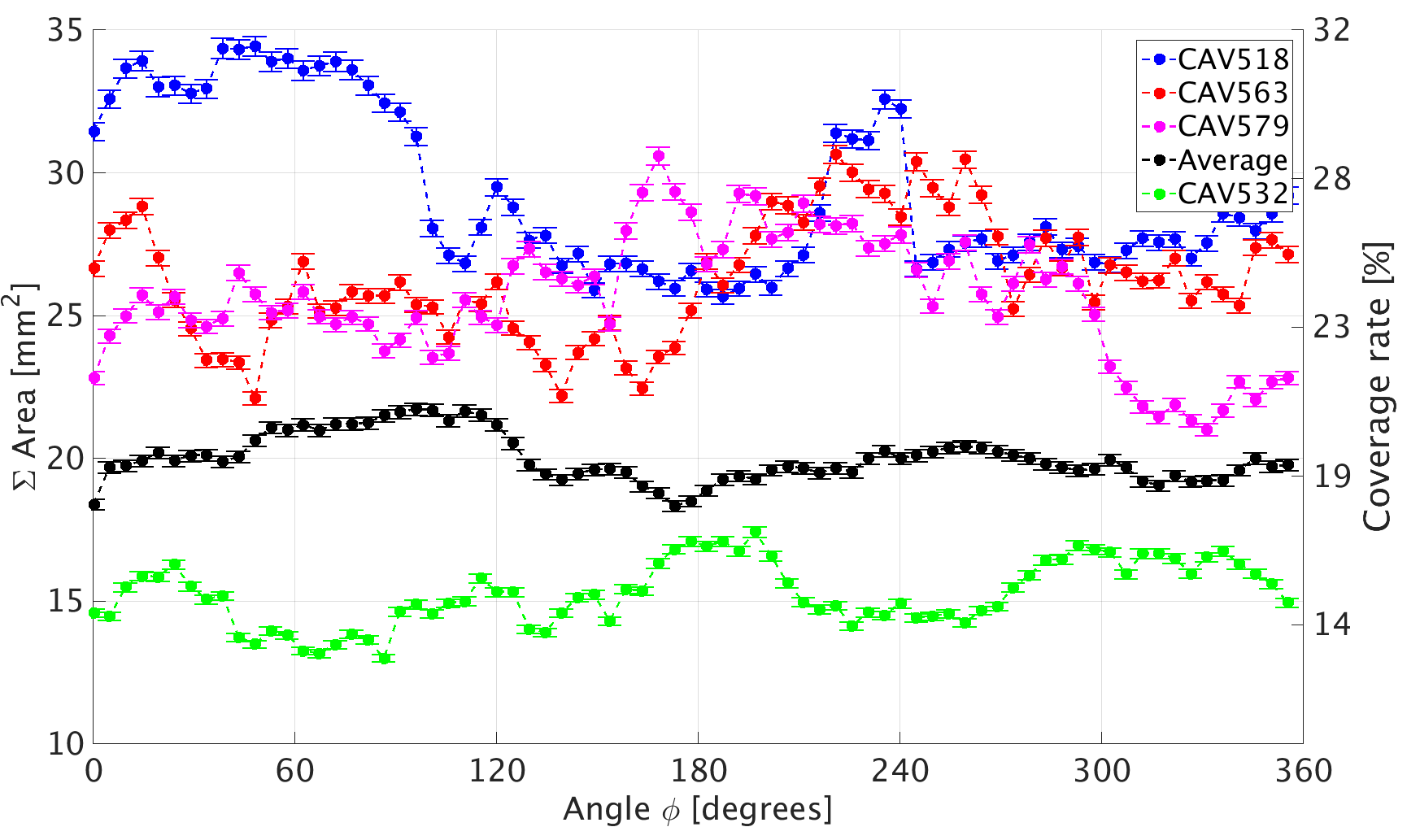}
   \caption{Integrated grain boundary area per image against the angular position of the image taken for the three cells with the largest values, the average of the whole set and the cell with the smallest value.}
   \label{Angular_worst_mean}
\end{figure}

\begin{figure}[!htb]
   \centering
   \vspace{5mm}
   \includegraphics*[width=0.50\textwidth]{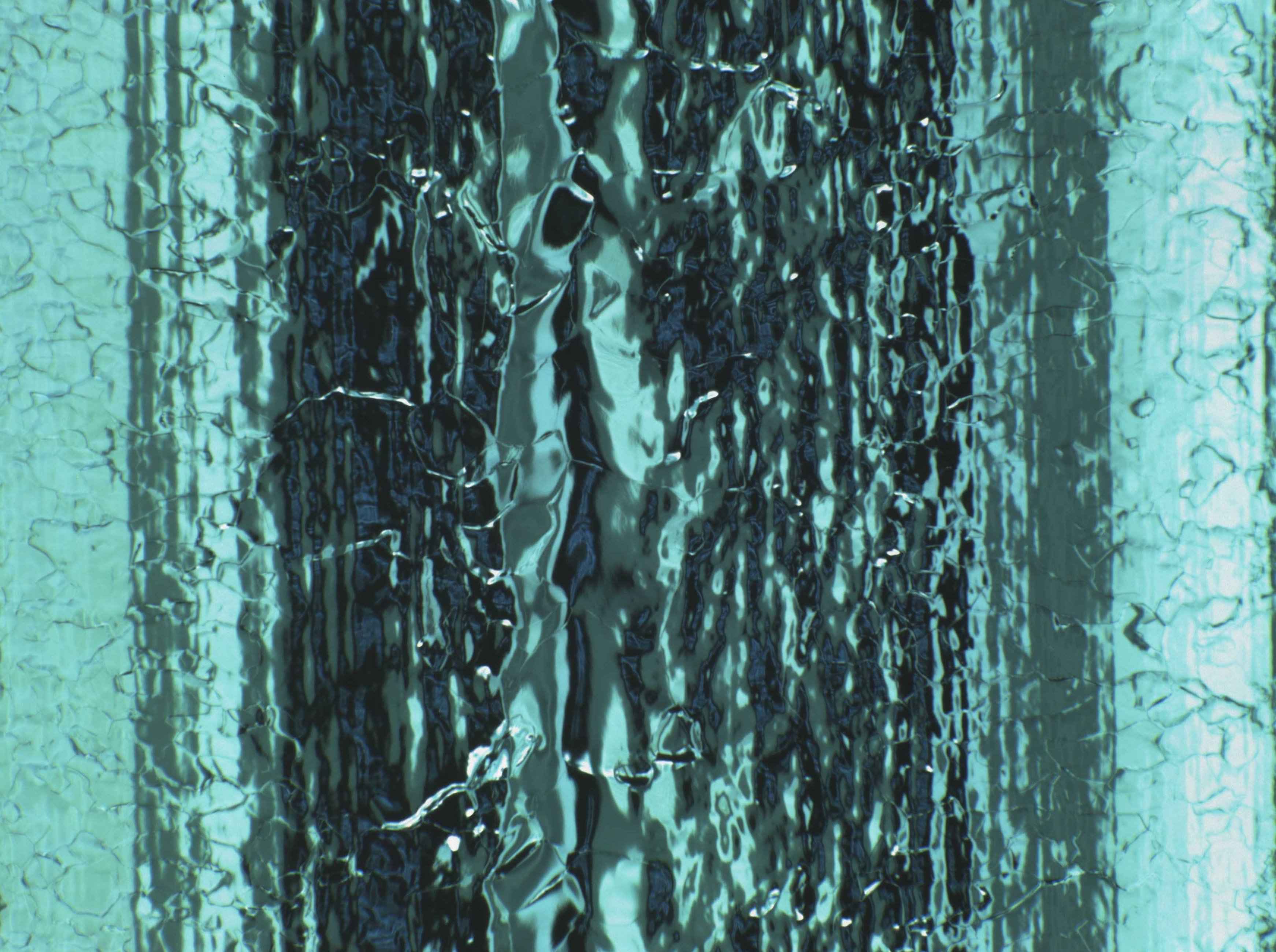}
   \caption{OBACHT image of the inner cavity surface of cavity CAV518, equator 9, 225.6 degree.}
   \label{CAV518}
\end{figure}

\begin{figure}[!htb]
   \centering
   \vspace{5mm}
   \includegraphics*[width=0.50\textwidth]{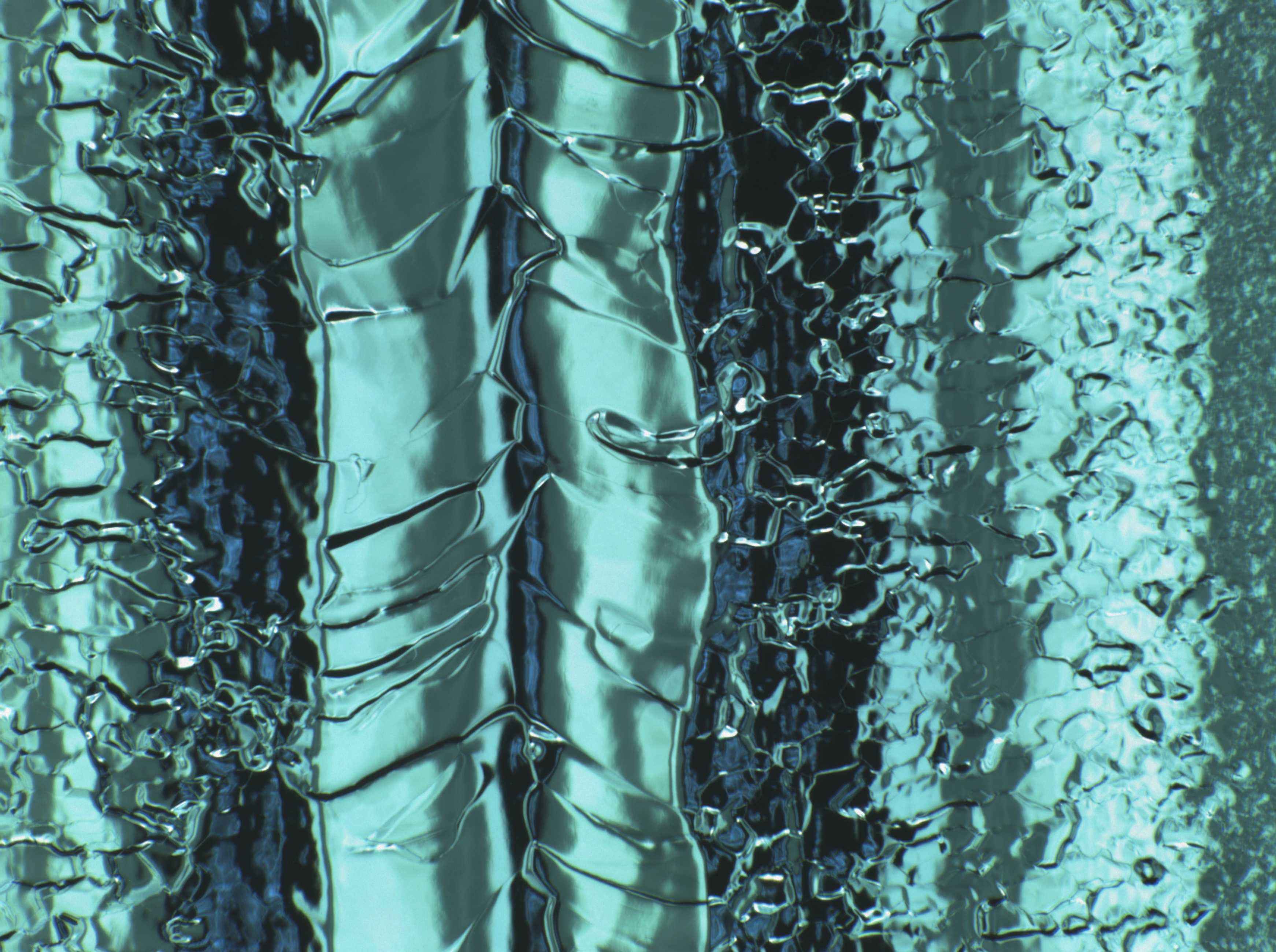}
   \caption{OBACHT image of the inner cavity surface of cavity CAV563, equator 9, 249.6 degree.}
   \label{CAV563}
\end{figure}

\begin{figure}[!htb]
   \centering
   \vspace{5mm}
   \includegraphics*[width=0.50\textwidth]{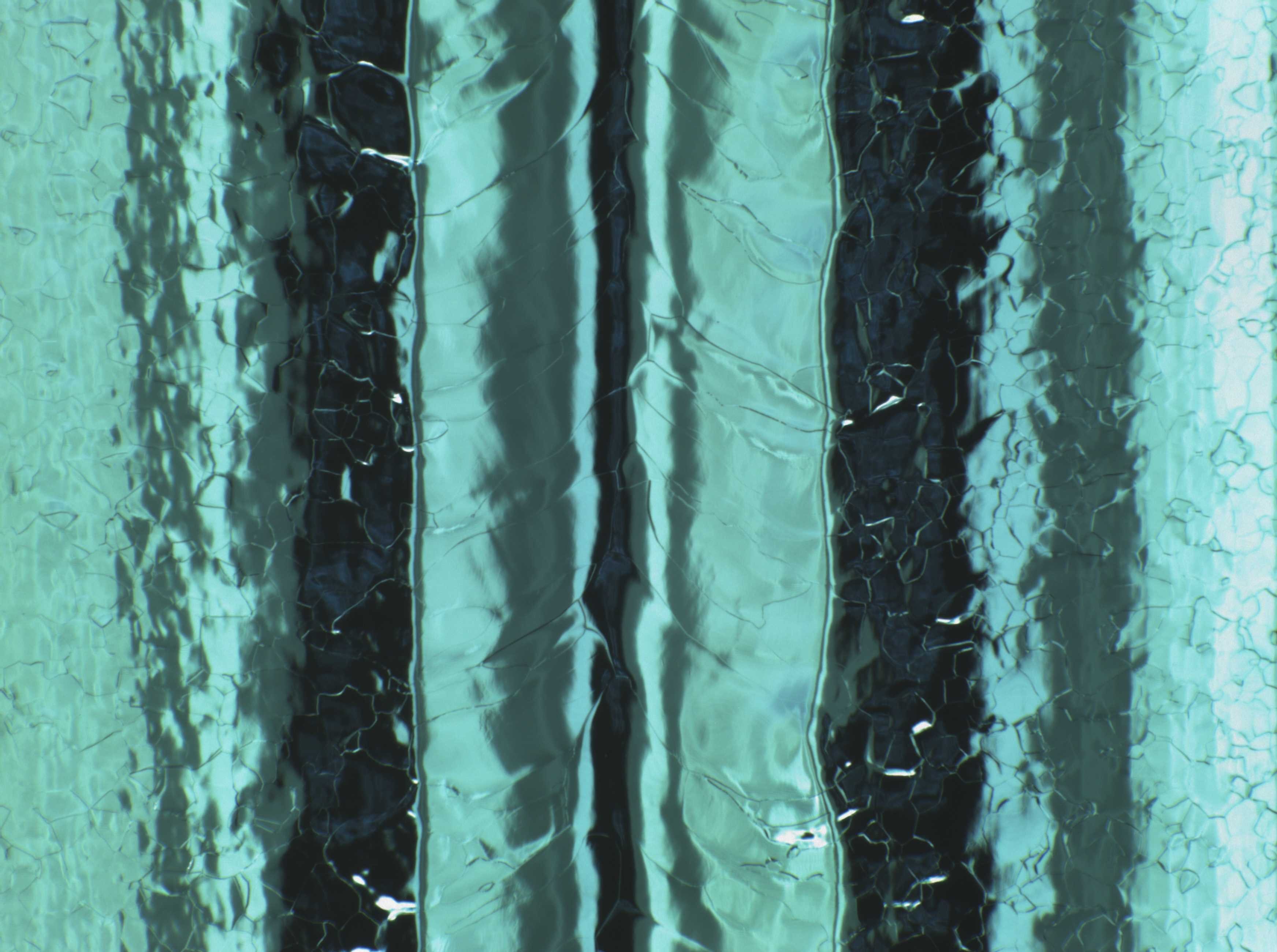}
   \caption{OBACHT image of the inner cavity surface of cavity CAV532, equator 6, 182.4 degree.}
   \label{CAV532}
\end{figure}

The integrated boundary area of three cells of EZ cavities exceed the average value by more than 25\,\% and one cell is below the average value by 25\,\%, see Figure \ref{Hist_EZ}. For a better understanding of the origin of this deviation, the integrated grain boundary area per image against the angular position of the image taken for these cells is shown in Figure \ref{Angular_worst_mean}.
Images of the specific cells significantly exceeding the average are shown in Figures \ref{CAV518} and \ref{CAV563}. In comparison to this excess, Figure \ref{CAV532} shows an image of the inner cavity surface with a smaller than average integrated grain boundary area per image.

CAV518 (Figure \ref{CAV518}) shows some remnants of the machining procedure prior to welding. Although all equators of CAV00518 showed a larger integrated grain boundary area than average, equator nine was outstanding. The same equator was identified as quench origin. 

CAV563 (Figure \ref{CAV563}) shows prominent grain boundaries in the welding seam and heat affected zone. 
This topography was observed in several cavities, as well as in CAV579. It was usually observed in equator nine
and in varying intensities and angular range. 

CAV532 (Figure \ref{CAV532}) is a cavity with an overall homogenous and smooth appearance. 
\newpage
In order to estimate the influence of the surface chemistry on the observed values of the integrated grain boundary area, images of the bulk niobium apart of the weld were analyzed. Those so called \textit{cell images} are taken with an offset of $\pm\,11.5\,\mathrm{mm}$ relative to the equator images during the inspections. 
Hence, only an influence of the surface chemistry and not from the electron beam welding in these images is expected. 
In contrast to the equator images, where a huge spread of the integrated grain boundary area was observed, the values for the inner surface in the cell are comparable within uncertainties for all inspected cavities of both vendors.

\subsection{Correlation with RF Performance}
\begin{figure}[!htb]
	\centering
		\includegraphics[width=1.00\textwidth]{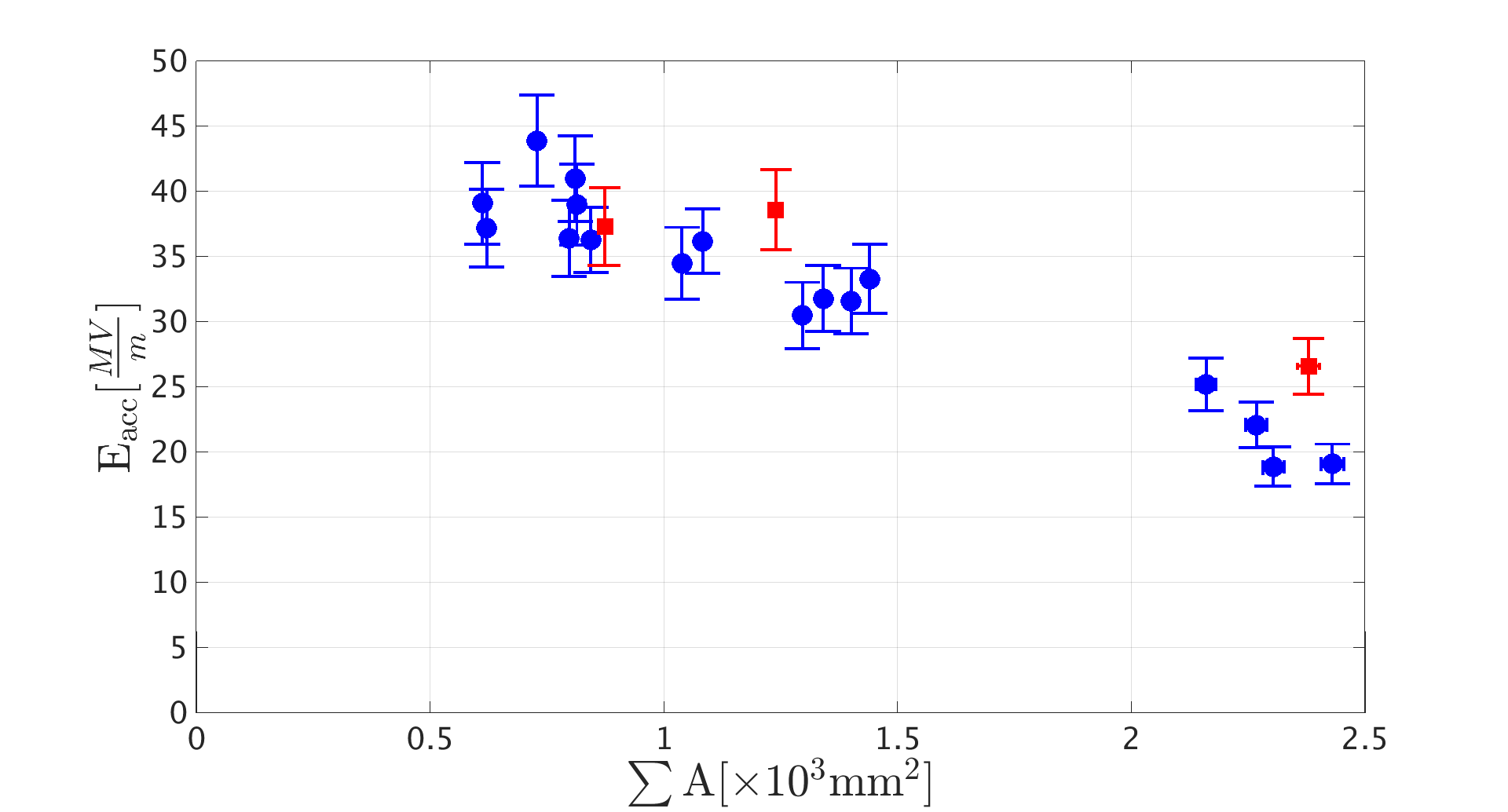}
	\caption{On the x-axis, the integrated grain boundary area of the optically conspicuous cell, which stands for a complete cavity, is shown, while the y-axis depicts the maximum accelerating field achieved by the respective cavity. The red squares display large grain cavities, the blue circles fine grain cavities.}
	\label{fig:Eacc_sumarea_wLG}   
\end{figure}

Table \ref{Values} shows a comparison of the achieved accelerating field of a cavity 
against the average integrated grain boundary area per image of the optically conspicuous cell of this cavity.  
\begin{table}[h!t]
	\centering
    \setlength\tabcolsep{9.5pt}
    \begin{tabular}{ccc}
        \toprule
        \textbf{Cavity} & \textbf{$E_{\mathrm{acc,max}}~ [\mathrm{MV/m}]$}               & \textbf{$<\upsigma \mathrm{A}>_{\mathrm{im}}~[\mathrm{mm^2}]$} \\
	\midrule
CAV518       & 22                 & 28  \\
CAV563       & 20                 & 26  \\
CAV579       & 25                 & 24  \\
\midrule
$<\mathrm{EZ}>$         & 28 $\pm$ 7         & 20 $\pm$ 2  \\
\midrule
CAV532       & 35                 & 15  \\
  \bottomrule
    \end{tabular}
		    \caption{Comparison of the RF results and the integrated grain boundary area per image.}
				    \label{Values}

\end{table}
For a more quantitatively approach, a larger set of cavities was chosen. In order not to be affected by local defects and to deduce an unbiased correlation between optical surface properties and the RF performance of a cavity, a set of cavities are selected by the following criteria:
\begin{enumerate}
	\item No surface chemistry between optical inspection and the cold RF test,
	\item Optical inspection shows no local defect,
	\item No field emission during the RF test. 
\end{enumerate}
The first criterion is needed to assure that the results of the two methods, optical inspection and cold RF test, can be correlated. A local defect, which is more likely to be the cause of a possible limitation of the cavity RF performance, has to be avoided in order to study the correlation between global surface properties and the RF performance. This is the reason for the second criterion. The last criterion prevents a falsification of the RF performance, because field emission is caused by a local defect in the highest electric field region, which is the iris region for TESLA type cavities, and introduces a different loss mechanism. A total number of 17 cavities from the XFEL production fulfill the before mentioned criteria, nine RI and eight EZ cavities. To increase the data set, but also to improve the universality of this study, three so-called large grain cavities were included, namely AC151, AC153 and AC154\footnote[1]{The grain size for XFEL cavities had to be on average ASTM 5 or 50\,$\upmu \mathrm{m}$ for the sheets before welding. Large grain cavities have grain sizes on the order of several centimeters.These cavities were produced at RI and underwent a main BCP with a removal of 100-110\,$\upmu \mathrm{m}$}. They fulfill the above mentioned criteria and increase the data set to a total number of 20 cavities.
Figure \ref{fig:Eacc_sumarea_wLG} shows the measured maximum accelerating field as a function of the integrated grain boundary area.
The correlation coefficients are given in Table \ref{tab:Eacc_sum_area}.
\begin{table}[ht]
\small
	\centering
\makebox[\linewidth]{
		\begin{tabular}{l>{\centering\arraybackslash} p{2cm}>{\centering\arraybackslash}    p{2cm}>{\centering\arraybackslash}  p{2cm}>{\centering\arraybackslash}  p{2cm}}
 \toprule
												& All  & RI & E.Z. & Sheets \\
								 \cmidrule{2-5}												
		\textbf{Pearson $\rho$}   	&  -0.93 & -0.91 & -0.96 & -0.97    \\
		\textbf{Corrected $\rho$}  		    &  -0.93 & -0.90 & -0.96 & -0.97    \\      
		\textbf{95\% c.i.}  &[-0.96,-0.84]&  [-0.99,-0.5] & [-0.99,-0.54] &  [-0.99,-0.82]  \\
		\textbf{Significance $\upsigma$}  & 6  & 3.4 & 4 & 5.2 \\
		\bottomrule
		\end{tabular}
		}
			\caption{Pearson correlation coefficients for the variables $\sum{\mathrm{A}}$ and $\mathrm{E_{acc,max}}$ for different subgroups.}
		\label{tab:Eacc_sum_area} 
\end{table}
\newline
A very strong negative correlation of $\rho\,=\,-0.93$ between the two variables $\sum{\mathrm{A}}$ and $\mathrm{E_{acc,max}}$ was found. The result is consistent for different subgroups, which were tested to reveal a systematic origin of the observed correlation. Those three subgroups are the (1) nine RI cavities, (2) the eight EZ cavities, and (3) eleven cavities from both vendors but from the same niobium supplier (Sheets). The large grain cavities are only included in the complete sample. The statistical significance of the correlation was found to be $6\,\upsigma$. The reduced significance in the subgroups is caused by the reduced dataset. 

\section{Discussion}
\begin{figure}[!htb]
	\centering
		\includegraphics[width=0.6\textwidth]{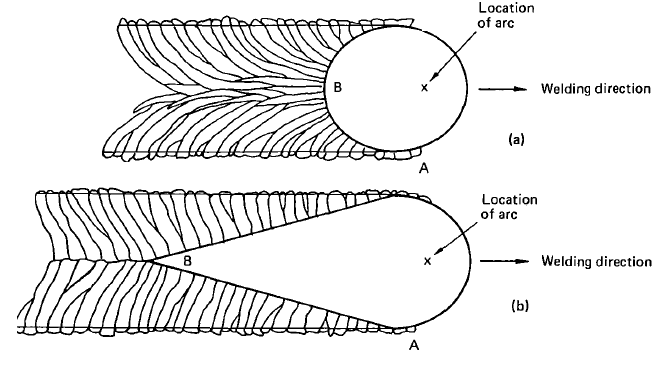}
	\caption{Comparison of welding puddle shapes. Travel speeds: (a)slow, (b) fast. Note that the orientation of the grains change during growth in (a) while the orientation is constant in (b). This is a consequence of the spatial orientation of the thermal gradient. While in (a) and (b) the thermal gradient at point A is $90^\mathrm{o}$ to the weld axis, in (a) the thermal gradient at B is parallel to the weld axis while it changes the orientation only slightly (b).}
	\label{WeldingFeed}
\end{figure}

The observed vendor specific grain boundary orientation as seen in Figure \ref{Orienthist_vendors} can be explained by the vendor specific EBW parameters. The travel speed influences the overall bead shape as the shape changes from elliptical to tear drop shaped as the welding speed increases (see Figure \ref{WeldingFeed} taken from \cite{EBW1997}). The grains will grow in the direction of the thermal gradient, starting at the base metal and into the liquified niobium.   

Comparing this with the observed pattern shows that RI has a beam travel speed slower than EZ, while the exact values are not public. But given the data and experience in \cite{Udomphol2007,EBWBuch}, it can be estimated that the welding speed for EZ has to be on the order of $16\,\pm\,1\,\frac{\mathrm{mm}}{\mathrm{s}}$ while for RI it has to be on the order of $5\,\pm\,1\,\frac{\mathrm{mm}}{\mathrm{s}}$. Those values are in good agreement with available data on welding speeds for different cavities and their grain boundary orientations \cite{Singer2017,Geng1999,Schmidt10}.

\begin{figure}[!htb]
	\centering
		\includegraphics[width=0.60\textwidth]{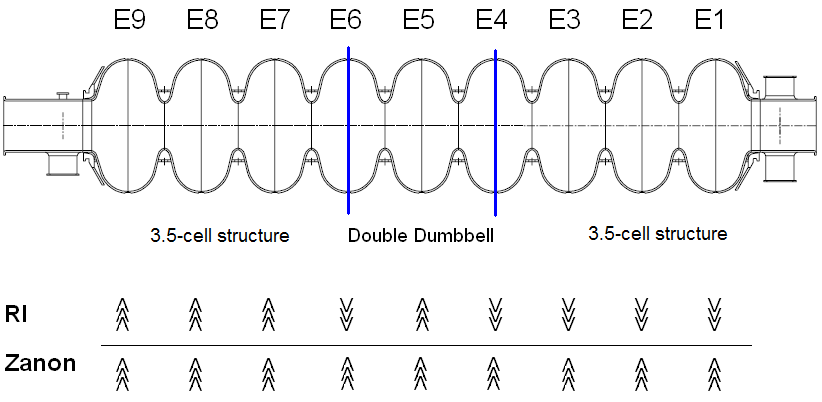}
	\caption{The upper schematic sketches the assembly procedure of a cavity at the two vendors and allows for the assignment of the patterns to the corresponding cells. The lower sketch depicts the orientation pattern as found in the respective cell.}
	\label{VendorAssembly}
\end{figure}

The specific series of welding patterns which are shown in Figure \ref{Orienthist_vendors_patterns} are identical for all cavities from a vendor and can be explained with the vendor specific assembly procedure (see Figure \ref{VendorAssembly}). 
While EZ assembles all dumbbells in one group and weld them in one EBW run - where the equators are welded subsequently in an alternating sequence - the assembling process for RI is split up in four steps. 

The first two steps are the assembly of two different structures consisting of three dumbbells and one end group, which are welded together ([E1-3] and [E7-9]),but performed separately. 

The third step is the weld of one double-dumbbell (E5). Between each step, the fabricated parts are brought out of the EBW machine. For the final step, the weld of equator 4 and 6, one end cell group is rotated by $180^\mathrm{o}$ with respect to the other. The different welding sequences result in a distinct welding pattern along the cavity for each vendor. A single image of the first equator can identify the cavity vendor.

The observed vendor-specific surface roughness (see Figure \ref{Vendor_Rdq_Distribution} and Table \ref{tab:Rdq_emg_fit_vendor} and \ref{tab:Rdq_emg_fit_vendor_mlv}) are in good agreement with the surface topology of EP- and BCP-treated cavities. It is known that EP leads to a smaller average roughness than BCP \cite{Padamsee2008,Ciovati2008}, as well as smaller boundary step heights and slopes \cite{Geng1999,Xu2011}, which is reflected in the average $\mathrm{R_{dq}}$. In addition, a spatial inhomogeneous removal of material by EP has been observed \cite{Geng1999}, where the welding seam regions are more affected than the heat affected zone due to the inhomogeneous electric field along the cavity axis. This effect is visible in the mean of the Gaussian component, $\upmu$. The BCP treated cavities show a uniform roughness value while the values for EP treated cavities differ for the regions.

The agreement between the quench location obtained with the 2nd sound system (see Figure \ref{fig:RMSEMap_Yegor}) and the image analysis algorithm in two of two tests supports the assumption that the \textit{integrated grain boundary area} $\sum{\mathrm{A}}$ can be used as a correlator with the RF performance. Although the underlying mechanism for this correlation is not known, it is plausible that grain boundaries influence the RF performance. Given current developments for high quality factor and high accelerating gradient treatments, the influence of grain boundaries are yet to be investigated, although first results in the serial production of nitrogen doped cavities show such a dependency \cite{Grassellino2013, Grassellino2017, TTC1}.

The observed non-uniform longitudinal position of the optically conspicuous cell (see figure \ref{Vendor_worst}) is unexpected. The most likely optically conspicuous cells are cells with a special position in the vendor specific welding sequence. For EZ, equator 9, which is the optically conspicuous cell in 24 of 31 cavities, is the first weld to be welded while the other welds follow in an alternating sequence. At RI, the first step is the fabrication of assembly groups in which the assembly and equatorial welds of two 3.5 cell cavities [E1-3] and [E7-9] and one double dumbbell [E5] are done. In a final assembly and welding step of those assembly groups, equators 4 and 6 are welded. Equators 6 and 7 are the optically conspicuous cells in 9 of 14 cavities and equator 4 the optically best cell in 6 of 14 cavities. 

To better understand this behavior, exemplary cavities were investigated further, namely CAV518, CAV563, CAV579, and CAV532. These specific welding seam surfaces show distinct grain boundary topographies. The inspection of the cell images, containing the bulk niobium with no influence of the welding procedure, showed no comparable topography nor a vendor specific surface structure, which should be the case if the surface chemistry alone is the cause for the deviations found. This leads to the conclusion that the assembly and electron beam welding procedure have the most significant influence on the integrated grain boundary area.

The observed correlation, quantified in Table \ref{tab:Eacc_sum_area} and shown in Figure \ref{fig:Eacc_sumarea_wLG}, suggest that higher $E_{\mathrm{acc,max}}$ can be reached when the integrated grain boundary area $\sum{\mathrm{A}}$ is small. As mentioned, the underlying mechanism of this correlation is not clear, especially since the mentioned models describe loss mechanisms and do not discuss achievable accelerating gradients. This is a weak point of this analysis, since it can only state a correlation. To improve this analysis another approach was to model the achieved maximal gradient as the result of the interplay of a given quality factor at low fields and loss mechanisms depending on the integrated grain boundary area $\sum{\mathrm{A}}$. This model showed no correlation and was rejected. The next steps will be to include the observed surface roughness $\mathrm{R_{dq}}$, the orientation of the grain boundaries, material parameters of the cavities, and other parametrization of the RF performance in the analysis. 

\section{Conclusion}
In the scope of this work, a framework has been developed which allows an automated analysis of images of the inner surface of SRF Tesla type cavities by means of the optical inspection robot OBACHT. A first application of the newly developed framework was an investigation of optical surface properties of the two cavity vendors for the European XFEL and significant differences in the quantitative characterization have been identified. Each difference turns out to be related to the vendor specific cavity assembly, electron beam welding and surface treatment procedures. In order to study the interplay between the optical surface properties and RF limitations, the concept of an optically conspicuous cell, which is assumed to be the limiting cell of the RF performance, was introduced. A crosscheck with 2nd sound tests of two cavities support the assumptions of the identity of the optically conspicuous cell and the limiting cell of the RF performance. A noteworthy observation has been made. A non-uniform distribution of the optically conspicuous cell was observed and in fact, this non-uniform distribution is vendor specific. This observation can be explained by the vendor specific assembly procedure and equatorial welding sequence. In addition, correlations of optical surface properties versus the maximal achievable accelerating field $\mathrm{E_{acc,max}}$ of 20 cavities has been investigated. A strong negative correlation of the integrated grain boundary area $\sum{\mathrm{A}}$ versus the maximal achievable accelerating field $\mathrm{E_{acc,max}}$ has been found. In conclusion, a quantitative analysis and characterization of a cavity surface by means of optical methods has been achieved, which can be adapted and used for the quality assurance of a future large scale cavity production.

\section{Acknowledgments}
I would like to thank B. van der Horst, J. Iversen, A. Matheisen, W.-D. M\"oller, D. Reschke and W. Singer (DESY) for their support, insights and valuable discussions. Furthermore, I would like to thank S. Aderhold (now FNAL), S. Karstensen (DESY), A. Navitski (now RI), J. Schaffran (DESY), L. Steder (DESY) and Y. Tamashevich (now HZB) for their work. Otherwise, mine would have been impossible. This work is funded from the EU 7th Framework Program (FP7/2007-2013) under grant agreement number 283745 (CRISP), "Construction of New Infrastructures - Preparatory Phase", ILC-HiGrade, contract number 206711, BMBF project  05H12GU9, and from the Alexander von Humboldt Foundation.

\end{document}